   \documentclass[10pt]{article}
   \usepackage[latin1]{inputenc}
   \usepackage{geometry}
   \usepackage{setspace}
   \usepackage{graphics} 
   \usepackage{verbatim}
   \usepackage{amsmath}
   \usepackage{url}
   \usepackage[authoryear, round, sort]{natbib}
  \usepackage{hyperref} 

  \title{Variable selection from random forests: application to gene expression data} 
   \author{\vspace{20pt}
     Ramón Díaz-Uriarte$^{1,3}$, Sara Alvarez de Andrés$^2$\\
   $¹$Bioinformatics Unit, $²$Cytogenetics Unit\\
   Biotechnology Programme\\
   Spanish National Cancer Center (CNIO)\\
   Melchor Fernández Almagro 3 \\
   Madrid, 28029\\
\vspace{20pt}
   Spain. \\
$^3$ Author for correspondence.\\
   \texttt{rdiaz@ligarto.org}\\
   \url{http://ligarto.org/rdiaz}\\
   }
   \date{
   \vspace*{40pt}
   2005-06-22 \\
\vspace{20pt}
   {\bf Running Head:} Gene selection with random forest.} 
  
\begin{document}
  \maketitle
  \newpage
  \begin{abstract}

  Random forest is a classification algorithm well suited for
  microarray data: it shows excellent performance even when most
  predictive variables are noise, can be used when the number of
  variables is much larger than the number of observations, and
  returns measures of variable importance. Thus, it is important to
  understand the performance of random forest with microarray data and
  its use for gene selection.

   We first show the effects of changes in parameters of random forest on the
   prediction error.  Then we present an approach for gene selection
   that uses measures of variable importance and error rate,
   and is targeted towards the selection of small sets of genes.  Using
   simulated and real microarray data, we show that the gene selection
   procedure yields small sets of genes while preserving predictive accuracy.

  We first show the effects of changes in parameters of random forest
  on the prediction error rate with microarray data. Then we present
  two approaches for gene selection with random forest: 1) comparing
  variable importance plots of variable importance from original and permuted data
  sets; 2) using backwards variable elimination. Using simulated and
  real microarray data, we show: 1) variable importance plots can be used to recover
  the full set of genes related to the outcome of interest, without
  being adversely affected by collinearities; 2) backwards variable
  elimination yields small sets of genes while preserving predictive
  accuracy (compared to several state-of-the art algorithms). Thus,
  both methods are useful for gene selection. 

  All code is available as an R package, varSelRF, from CRAN
\href{http://cran.r-project.org/src/contrib/PACKAGES.html}
{http://cran.r-project.org/src/contrib/PACKAGES.html} or from the supplementary
material page.

Supplementary information:
\href{http://ligarto.org/rdiaz/Papers/rfVS/randomForestVarSel.html}{http://ligarto.org/rdiaz/Papers/rfVS/randomForestVarSel.html}

  \end{abstract}

 \footnotetext[1]{To whom correspondence should be addressed}

\section{Introduction}

Random forest is an algorithm for classification developed by Leo
Breiman \citep{breiman-rf} that uses an ensemble of classification
trees \citep{cart, ripley-96, htf-01}. Each of the classification
trees is built using a bootstrap sample of the data, and at each split
the candidate set of variables is a random subset of the variables.
Thus, random forest uses both bagging (bootstrap aggregation), a
successful approach for combining unstable learners
\citep{breiman-bagging, htf-01}, and random variable selection for
tree building.  Each tree is unpruned (grown fully), so as to obtain
low-bias trees; at the same time, bagging and random variable
selection result in low correlation of the individual trees.  The
algorithm yields an ensemble that can achieve both low bias and low
variance (from averaging over a large ensemble of low-bias,
high-variance but low correlation trees).

Random forest has excellent performance in classification tasks,
comparable to support vector machines. Although random forest is not
widely used in the microarray literature \citep[but see][]{SaraRF,
  Izmir2004, Wu.Zhao2003, Gunther.Heyes2003, Man-rf,
  Schwender.Bolt2004}, it has several characteristics that make it
ideal for these data sets: a) can be used when there are many more
variables than observations; b) has good predictive performance
even when most predictive variables are noise; c) does not
overfit; d) can handle a mixture of categorical and continuous
predictors; e) incorporates interactions among predictor variables; f)
the output is invariant to monotone transformations of the predictors;
g) there are high quality and free implementations: the original
Fortran code from L.\ Breiman and A.\ Cutler, and an R package from
A.\ Liaw and M.\ Wiener \citep{rf-rnews}; h) there is little need to
fine-tune parameters to achieve excellent performance; i) returns measures of
variable (gene) importance. The most
important parameter to choose is $mtry$, the number of input variables
tried at each split, but it has been reported that the default value
is often a good choice \citep{rf-rnews}.  In addition, the user needs
to decide how many trees to grow for each forest ($ntree$) as well as
the minimum size of the terminal nodes ($nodesize$). These three
parameters will be throughly examined in this paper.

Given these promising features, it is important to understand the
performance of random forest compared to alternative state-of-the-art
prediction methods with microarray data, as well as the effects
of changes in the parameters of random forest. In this paper we present, as necessary
background for the main topic of the paper (gene selection), the first
through examination of these issues, including evaluating the effects
of $mtry$, $ntree$ and $nodesize$ on error rate using
nine real microarray data sets and simulated data.

The main question addressed in this paper is gene selection using random
forest.  A few authors have previously used variable selection with random
forest.  \citet{dudoit-inbook} and \citet{Wu.Zhao2003} use filtering approaches
and, thus, do not take advantage of the measures of variable importance
returned by random forest as part of the algorithm. \citet{svetnik} propose a
method that is somewhat similar to our approach. The main difference is that
\citet{svetnik} first find the ``best'' dimension ($p$) of the model, and then
choose the $p$ most important variables. This is a sound strategy when the
objective is to build accurate predictors, without any regards for model
interpretability.  But this might not be the most appropriate for our purposes
as it shifts the emphasis away from selection of specific genes, and in genomic
studies the identity of the selected genes is relevant (e.g., to understand
molecular pathways or to find targets for drug development).

The last issue addressed in this paper is the multiplicity (or lack of
uniqueness or lack of stability) problem. Variable selection with microarray
data can lead to many solutions that are equally good from the point of view of
prediction rates, but that share few common genes.  This multiplicity problem
has been emphasized by \citet{Somorjai2003} and recent examples are shown in
\citet{EinDor} and \citet{Michielis}. Although multiplicity of results is not a problem when
the only objective of our method is prediction, it casts serious doubts on the
biological interpretability of the results \citep{Somorjai2003}.  Unfortunately
most ``methods papers'' in bioinformatics do not evaluate the stability of the
results obtained, leading to a false sense of trust on the biological
interpretability of the output obtained. Our paper presents a through and
critical evaluation of the stability of the lists of selected genes with the
proposed (and two competing) methods.

\section{Variable selection methods}

\subsection{Two objectives of variable selection}

When facing gene selection problems, biomedical researchers often show
interest in one of the following objectives:

\begin{enumerate}
\item To identify relevant genes for subsequent research; this
  involves obtaining a (probably large) set of genes that are related
  to the outcome of interest, and this set should include genes even if they
  perform similar functions and are highly correlated.
  
\item To identify small sets of genes to be used for diagnostic
  purposes in clinical practice; this involves obtaining the smallest
  possible set of genes that can still achieve good predictive
  performance (thus, ``redundant'' genes should not be selected). 
\end{enumerate}

We will focus on the second objective. The use of random forest for the first
objective is under investigation and will be reported elsewhere.

\subsection{Variable importance from random forest}

Random forest returns several measures of variable importance. The
most reliable measure is based on the decrease
of classification accuracy when values of a variable in a node of a
tree are permuted randomly \citep{breiman-rf, Bureau2003},
and this is the measure of variable importance (in its unscaled
version ---see supplementary material) that we will use in the rest of
the paper.


\subsection{Backwards elimination of variables (genes) using OOB error}

To select gebes we can iteratively fit random forests, at each iteration
building a new forest after discarding those variables (genes) with the
smallest variable importances; the selected set of genes is the one that yields the
smallest error rate.  Random forest returns a measure of error rate based on
the out-of-bag cases for each fitted tree, the OOB error, and this is the
measure of error we will use.  Note that in this section we are using OOB
error to choose the final set of genes, not to obtain unbiased estimates of the
error rate of this rule.  Because of the iterative approach, the OOB error is
biased down and cannot be used to asses the overall error rate of the approach,
for reasons analogous to those leading to ``selection bias'' \citep{ambroise, simon-03}. To assess prediction error rates we will use the bootstrap, not
OOB error (see section \ref{boot}). (Using error rates
  affected by selection bias to select the optimal number of genes is
  not necessarily a bad procedure from the point of view of selecting
  the final number of genes; see \citet{Braga-Neto.Carroll2004}).

In our algorithm we examine all forests that result from eliminating,
iteratively, a fraction, $fraction.dropped$, of the genes (the
least important ones) used in the previous iteration. By default,
$fraction.dropped = 0.2$ which allows for relatively fast operation,
is coherent with the idea of an ``aggressive variable selection''
approach, and increases the resolution as the number of genes
considered becomes smaller.  We do not recalculate variable
importances at each step as \citet{svetnik} mention severe overfitting
resulting from recalculating variable importances. After fitting all
forests, we examine the OOB error rates from all the fitted random
forests. We choose the solution with the smallest number of genes
whose error rate is within $u$ standard errors of the minimum error
rate of all forests. 
Setting $u = 0$ is the same as selecting the set of genes that
leads to the smallest error rate.  Setting $u = 1$ is similar to the
common ``1 s.e.  rule'', used in the classification trees literature
\citep{ripley-96, cart}; this strategy can lead to solutions with
fewer genes than selecting the solution with the smallest error
rate, while achieving an error rate that is not
different, within sampling error, from the ``best solution''. In this
paper we will examine both the ``1 s.e. rule'' and the ``0 s.e.
rule''.




\section{Evaluation of performance}

\subsection{Data sets}
We have used both simulated and real microarray data sets to evaluate
the variable selection procedure. For the real
data sets, original reference paper and main features are shown in
Table \ref{datasets}. Further details are provided in the
supplementary material.

\begin{table}
\caption{\label{datasets} Main characteristics of the microarray data
  sets used.}
{\footnotesize
\begin{tabular}{l|lrrr}
Dataset & Original ref.&Genes&Patients&Classes \\
\hline
Leukemia &\citet{golub}&3051&38&2\\
Breast &\citet{vveer}&4869&78&2\\
Breast &\citet{vveer}&4869&96&3\\
NCI 60 &\citet{ross}&5244&61&8\\
Adenocar-\\
cinoma &\citet{ramas-03}&9868&76&2\\
Brain &\citet{pomeroy}&5597&42&5\\
Colon &\citet{alon}&2000&62&2\\
Lymphoma &\citet{alizadeh}&4026&62&3\\
Prostate &\citet{singh}&6033&102&2\\
Srbct &\citet{khan}&2308&63&4\\
\hline
\end{tabular}
}
\end{table}


To evaluate if the proposed procedure can recover the signal in the
data, we need to use simulated data, so that we know exactly which
genes are relevant.  Data have been simulated using different numbers
of classes of patients (2 to 4), number of independent dimensions (1
to 3), and number of genes per dimension (5, 20, 100).  In all cases,
we have set to 25 the number of subjects per class. Each independent
dimension has the same relevance for discrimination of the classes.
The data come from a multivariate normal distribution with variance of
1, a (within-class) correlation among genes within dimension of
0.9, and a within-class correlation of 0 between genes from different
dimensions, as those are independent.  The multivariate means have
been set so that the unconditional prediction error rate
\citep{mclach-dlda} of a linear discriminant analysis using one gene
from each dimension is approximately 5\%.  To each data set we have
added 2000 random normal variates (mean 0, variance 1) and 2000 random
uniform $[-1, 1]$ variates.  In addition, we have generated data sets
for 2, 3, and 4 classes where no genes have signal (all 4000 genes are
random).  For the non-signal data sets we have generated four
replicate data sets for each level of number of classes. Further
details are provided in the supplementary material.

\subsection{Competing methods}

We have compared the predictive performance of the variable selection
approach with: a) random forest without any variable selection (using
$mtry = \sqrt{number\ of \ genes}$, $ntree = 5000$, $nodesize =
1$); b) three other methods that have shown good
performance in reviews of classification methods with microarray data
\citep{dudoit-dlda, romualdi-03, bag-boost} but that do not include
any variable selection; c) two methods that carry out
variable selection. 

For the three methods that do not carry out variable selection,
\textbf{Diagonal Linear Discriminant Analysis (DLDA)}, \textbf{K
  nearest neighbor (KNN)}, and \textbf{Support Vector Machines (SVM)}
with linear kernel, we have used, based on \cite{dudoit-dlda}, the 200
genes with the largest $F$-ratio of between to within groups sums of
squares. For \textbf{KNN}, the number of neighbors ($K$) was
chosen by cross-validation as in \cite{dudoit-dlda}.

One of the methods that incorporates gene selection is
\textbf{Shrunken centroids (SC)}, developed by \cite{shrunkenc}. We
have used two different approaches to determine the best number of
features. In the first one, \textbf{SC.l}, we choose the number of
genes that minimizes the cross-validated error rate and, in case of
several solutions with minimal error rates, we choose the one with
largest likelihood. In the second approach, \textbf{SC.s}, we choose
the number of genes that minimizes the cross-validated error rate and,
in case of several solutions with minimal error rates, we choose the
one with smallest number of genes (larger penalty). The second method
that incorporates gene selection is \textbf{Nearest neighbor +
  variable selection (NN.vs)}, where we filter genes using the
F-ratio, and select the number of genes that leads to the smallest
error rate; in our implementation, we run a Nearest Neighbor
classifier (KNN with K = 1) on all subsets of genes that result from
eliminating $20\%$ of the genes (the ones with the smallest F-ratio)
used in the previous iteration.  This approach, in its many variants
(changing both the classifier and the ordering criterion) is popular
in microarray papers; a recent example is \cite{roepman}, and
similar general strategies are implemented in the program Tnasas
\citep{gepas2}. Further
details of all these methods are provided in the supplementary
material. All simulations and analyses were carried out with R
\citep[http://www.r-project.org; ][]{R}, using
packages randomForest (from A.\ Liaw and M.\ Wiener) for random
forest, e1071 (E.\ Dimitriadou, K.\ Hornik, F.\ Leisch, D.\ Meyer, and
A.\ Weingessel) for SVM, class (B.\ Ripley and W.\ Venables) for KNN,
PAM \citep{shrunkenc} for shrunken centroids, and
geSignatures (by R.D.-U.) for DLDA.

\subsection{\label{boot}Estimation of error rates} 
To estimate the prediction error rate of all methods we have used the
.632+ bootstrap method \citep{ambroise, 632-rule}. It must be
emphasized that the error rate used when performing variable selection
is not the error rate reported as the prediction error rate (e.g.,
Table \ref{error.rates}), nor the error used to compute the .632+
estimate. To calculate the prediction error rate (as reported, for
example, in Table \ref{error.rates}) the .632+ bootstrap method is
applied to the complete procedure, and thus the ``out-of-bag'' samples
used in the .632+ method are samples that are not used when fitting
the random forest, or carrying out variable selection. This also
applies when evaluating the competing methods.

\subsection{Stability (uniqueness) of results}
Following \citet{Faraway-92}, \citet{harrell-01}, and
\citet{efron-gong},  we have evaluated
the stability of the variable selection procedure using the
bootstrap.  This allows us to asses how often a given
gene, selected when running the variable selection procedure in the
original sample, is selected when running the procedure on bootstrap
samples.

 \begin{figure}                                                         
 \begin{center}
 {\resizebox{!}{7.5cm}{%
 \includegraphics{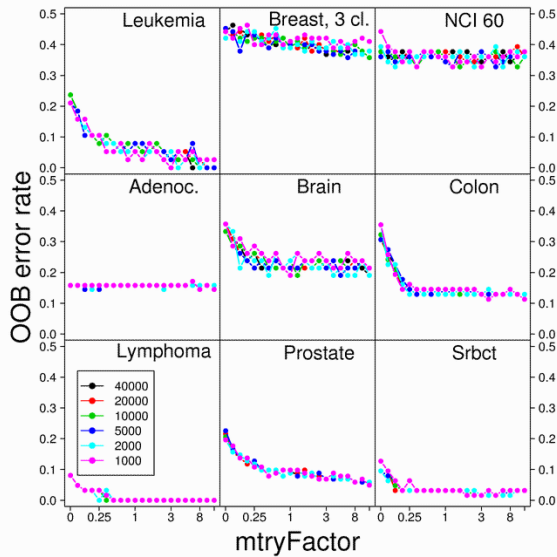}}}


\caption{\label{mtry.ntree.paper.real} Out-of-Bag (OOB) vs
  $mtryFactor$ for the nine microarray data sets.  $mtryFactor$ is the
  multiplicative factor of the default $mtry$
  ($\sqrt{number.of.genes}$); thus, an $mtryFactor$ of 3 means the
  number of genes tried at each split is $3 *\sqrt{number.of.genes}$;
  an $mtryFactor = 0$ means the number of genes tried was 1; the
  $mtryFactor$s examined were $= \{0, 0.05, 0.1, 0.17, 0.25, 0.33, 0.5,
  0.75, 0.8, 1, 1.15, 1.33, 1.5, 2, 3,$ $4, 5, 6, 8, 10, 13\}$. Results
  shown for six different $ntree = \{1000, 2000, 5000,
  10000, 20000, 40000\}$.  $nodesize = 1$.}
\end{center}
\end{figure}

\section{Results}

 \begin{table*}[b!]  \begin{center} 
       \caption{\label{error.rates} Error rates (estimated using the
         0.632+ bootstrap method with 200 bootstrap samples) for the
         microarray data sets using different methods (see text for
         description of alternative methods).  The results shown for
         variable selection with random forest used $ntree = 2000,
         fraction.dropped = 0.2, mtryFactor = 1$.  Note that the OOB
         error used for variable selection \emph{is not} the error
         reported in this table; the error rate reported is obtained
         using bootstrap on the complete variable selection process.
         The column ``no info'' denotes the minimal error we can make
         if we use no information from the genes (i.e., we always bet
         on the most frequent class).}

  {\footnotesize
    \begin{tabular}{l|cccccccccc}

Data set& no info & SVM & KNN & DLDA& SC.l & SC.s & NN.vs & random forest &
\multicolumn{2}{c}{random forest var.sel.}\\
&& & & & & & & & s.e.\ 0 & s.e.\ 1\\

\hline
Leukemia &       0.289 &0.014 &  0.029 &  0.020 &   0.025& 0.062  &   0.056&      0.051 &   0.087  &   0.075   \\
Breast 2 cl.&    0.429 &0.325 &  0.337 &  0.331 &   0.324& 0.326  &  0.337&       0.342 &   0.337  &   0.332  \\
Breast 3 cl.&    0.537 &0.380 &  0.449 &  0.370 &   0.396& 0.401  &   0.424&      0.351 &   0.346  &   0.364  \\
NCI 60      &    0.852 &0.256 &  0.317 &  0.286 &   0.256& 0.246  &   0.237&      0.252 &   0.327  &   0.353  \\
Adenocar.&       0.158 &0.203 &  0.174 &  0.194 &   0.177 & 0.179 &    0.181&     0.125 &   0.185  &   0.207  \\
Brain&           0.761 &0.138 &  0.174 &  0.183 &   0.163 & 0.159 &    0.194&     0.154 &   0.216  &   0.216  \\
Colon&           0.355 &0.147 &  0.152 &  0.137 &   0.123 & 0.122 &    0.158&     0.127 &   0.159  &   0.177  \\
Lymphoma &       0.323 &0.010 &  0.008 &  0.021 &   0.028 & 0.033 &    0.04 &     0.009 &   0.047  &   0.042  \\
Prostate &       0.490 &0.064 &  0.100 &  0.149 &   0.088 & 0.089 &    0.081&     0.077 &   0.061  &   0.064  \\
Srbct &          0.635 &0.017 &  0.023 &  0.011 &   0.012 & 0.025 &    0.031&     0.021 &   0.039  &   0.038   \\
\hline
\end{tabular}
}
\end{center}
\end{table*}

\subsection{Choosing $mtry$ and $ntree$}

Preliminary data suggested that $mtry$ and $ntree$ could affect the shape of
variable importance plots.  At the same time, use of OOB error rate as a
guidance to select $mtry$ could be affected by $ntree$ and, potentially,
$nodesize$. Thus, we first examined whether the OOB error rate is substantially
affected by changes in $mtry$, $ntree$, and $nodesize$.

Figure \ref{mtry.ntree.paper.real} and the supplementary material (Figure
\\``error.vs.mtry.pdf''), however, show that, for both real and simulated data,
the relation of OOB error rate with $mtry$ is largely independent of $ntree$
(for $ntree$ between 1000 and 40000) and $nodesize$ (nodesizes 1 and 5). In
addition, the default setting of $mtry$ ($mtryFactor = 1$ in the figures) is
often a good choice in terms of OOB error rate. In some cases, increasing
$mtry$ can lead to small decreases in error rate, and decreases in $mtry$ often
lead to increases in the error rate. This is specially the case with simulated
data with very few relevant genes (with very few relevant genes, small $mtry$
results in many trees being built that do not incorporate any of the relevant
genes). Since the OOB error and the relation between OOB error and $mtry$ do
not change whether we use $nodesize$ of 1 or 5, and because the increase in
speed from using $nodesize$ of 5 is inconsequential, all further analyses will
use only the default $nodesize = 1$.

 \subsection{Backwards elimination of variables (genes) using OOB
   error} On the simulated data sets (see supplementary material,
 Tables 3 and 4) 
 backwards elimination often leads to very small sets of genes, often
 much smaller than the set of ``true genes''. The error rate of the
 variable selection procedure, estimated using the .632+ bootstrap
 method, indicates that the variable selection procedure does not lead
 to overfitting, and can achieve the objective of aggressively
 reducing the set of selected genes.  In contrast, when the
 simplification procedure is applied to simulated data sets without
 signal (see Tables 1 and 2 
in supplementary material), the number of
 genes selected is consistently much larger and, as should be the
 case, the estimated error rate using the bootstrap corresponds to
 that achieved by always betting on the most probable class.

Results for the real data sets are shown in Tables \ref{error.rates} and
\ref{stability} (see also supplementary material, Tables 5, 6, 7, 
for additional results using different combinations of $ntree =
\{2000,5000,20000\}$, $mtryFactor = \{1, 13\}, se=\{0, 1\},
fraction.dropped=\{0.2, 0.5\}$). Error rates (see Table
\ref{error.rates}) when performing variable selection are in most cases comparable
(within sampling error) to those from random forest without variable
selection, and comparable also to the error rates from competing
state-of-the-art prediction methods. The number of genes selected
varies by data set, but generally (Table \ref{stability}) the
variable selection procedure leads to small ($< 50$) sets of predictor
genes, often much smaller than those from competing approaches
(see also Table 8 in supplementary material). There are no relevant
differences in error rate related to differences in $mtry$, $ntree$ or
whether we use the ``s.e.\ 1'' or ``s.e.\ 0'' rules. The use of the
``s.e.\ 1'' rule, however, tends to result in smaller sets of selected
genes.

\subsection{Stability (uniqueness) of results}
The results here will focus on the real microarray data sets (results
from the simulated data are presented on the supplementary material).
Table \ref{stability} (see also supplementary material, Tables 5, 6, 7,
for other combinations of $ntree, mtryFactor, fraction.dropped, se$)
shows the variation in the number of genes selected in bootstrap
samples, and the frequency with which the genes selected in the
original sample appear among the genes selected from the bootstrap
samples. In most cases, there is a wide range in the number of genes
selected; more importantly, the genes selected in the original samples
are rarely selected in more than 50\% of the bootstrap samples. These
results are not strongly affected by variations in $ntree$ or $mtry$;
using the ``s.e.\ 1'' rule can lead, in some cases, to increased
stability of the results.

As a comparison, we also show in Table \ref{stability} the stability
of two alternative approaches for gene selection, the shrunken
centroids method, and a filter approach combined with a Nearest
Neighbor classifier (see Table 8 in the supplementary material for
results of SC.l). Error rates are comparable, but both alternative
methods lead to much larger sets of selected genes than backwards
variable selection with random forests. The alternative approaches
seem to lead to somewhat more stable results in variable selection (probably a
consequence of the large number of genes selected) but
in practical applications this increase in stability is probably far
out-weighted by the very large number of selected genes.

 \begin{table}[p]
 \begin{center}
   \caption{\label{stability} Stability of variable (gene) selection evaluated
     using 200 bootstrap samples. ``\# Genes'': number of genes
     selected on the original data set. ``\# Genes boot.'': median
     (1st quartile, 3rd quartile) of number of genes selected from 
     on the bootstrap samples. ``Freq. genes'': median (1st quartile,
     3rd quartile) of the frequency with which each gene in the
     original data set appears in the genes selected from the
     bootstrap samples. Parameters for backwards elimination with
     random forest: $mtryFactor = 1, s.e.\ = 0, ntree = 2000,
     ntreeIterat = 1000, fraction.dropped = 0.2$.}
 \end{center}
 \begin{center}
\vspace{-32pt} 
 {\footnotesize
 \begin{tabular}{l|rrrr}
 Data set& Error & \# Genes & \# Genes boot. & Freq. genes\\
 \hline
 \hline
 \multicolumn{5}{c}{\textbf{Backwards elimination of genes from random forest}}\\ 
 \hline
\multicolumn{5}{c}{$s.e.\ = 0$}\\ 
 \hline
 Leukemia    &  0.087 &   2 &  2 (2, 2) &   0.38 (0.29, 0.48)\footnotemark[1]\\ 
 Breast 2 cl.&  0.337 &  14 &  9 (5, 23)&   0.15 (0.1, 0.28)\\ 
 Breast 3 cl.&  0.346 & 110 & 14 (9, 31)&   0.08 (0.04, 0.13)\\ 
 NCI 60      &  0.327 & 230 & 60 (30, 94)&    0.1 (0.06, 0.19)\\ 
 Adenocar.   &  0.185 &   6 &  3 (2, 8)&   0.14 (0.12, 0.15)\\ 
 Brain       &  0.216 &  22 & 14 (7, 22)&   0.18 (0.09, 0.25)\\ 
 Colon       &  0.159 &  14 &  5 (3, 12)&   0.29 (0.19, 0.42)\\ 
 Lymphoma    &  0.047 &  73 & 14 (4, 58)&   0.26 (0.18, 0.38)\\ 
 Prostate    &  0.061 &  18 &  5 (3, 14)&   0.22 (0.17, 0.43)\\ 
 Srbct       &  0.039 & 101 & 18 (11, 27)&    0.1 (0.04, 0.29)\\ 
 \hline
 \hline
\multicolumn{5}{c}{$s.e.\ = 1$}\\ 
 \hline
 Leukemia    & 0.075 &  2 &  2 (2, 2)&    0.4 (0.32, 0.5)\footnotemark[1]\\ 
 Breast 2 cl.& 0.332 & 14 &  4 (2, 7)&   0.12 (0.07, 0.17)\\ 
 Breast 3 cl.& 0.364 &  6 &  7 (4, 14)&   0.27 (0.22, 0.31)\\ 
 NCI 60      & 0.353 & 24 & 30 (19, 60)&   0.26 (0.17, 0.38)\\ 
 Adenocar.   & 0.207 &  8 &  3 (2, 5)&   0.06 (0.03, 0.12)\\ 
 Brain       & 0.216 &  9 & 14 (7, 22)&   0.26 (0.14, 0.46)\\ 
 Colon       & 0.177 &  3 &  3 (2, 6)&   0.36 (0.32, 0.36)\\ 
 Lymphoma    & 0.042 & 58 & 12 (5, 73)&   0.32 (0.24, 0.42)\\ 
 Prostate    & 0.064 &  2 &  3 (2, 5)&    0.9 (0.82, 0.99)\footnotemark[1]\\ 
 Srbct       & 0.038 & 22 & 18 (11, 34)&   0.57 (0.4, 0.88)\\ 
 \hline
 \hline
 \multicolumn{5}{c}{\textbf{Alternative approaches}}\\ 
 \hline
  \multicolumn{5}{c}{SC.s}\\
 \hline
 Leukemia    & 0.062 &   82\footnotemark[2] &   46 (14, 504)&   0.48 (0.45, 0.59)\\ 
 Breast 2 cl.& 0.326 &   31 &   55 (24, 296)&   0.54 (0.51, 0.66)\\ 
 Breast 3 cl.& 0.401 & 2166 & 4341 (2379, 4804)&   0.84 (0.78, 0.88)\\ 
 NCI 60      & 0.246 & 5118 & 4919 (3711, 5243)&   0.84 (0.74, 0.92)\\ 
 Adenocar.   & 0.179 &    0 &    9 (0, 18)&     NA (NA, NA)\\ 
 Brain       & 0.159 & 4177 & 1257 (295, 3483)&   0.38 (0.3, 0.5)\\ 
 Colon       & 0.122 &   15 &   22 (15, 34)&    0.8 (0.66, 0.87)\\ 
 Lymphoma    & 0.033 & 2796 & 2718 (2030, 3269)&   0.82 (0.68, 0.86)\\ 
 Prostate    & 0.089 &    4 &    3 (2, 4)&   0.72 (0.49, 0.92)\\ 
 Srbct       & 0.025 &   37\footnotemark[3] &   18 (12, 40)&   0.45 (0.34, 0.61)\\ 
 \hline
 \hline
\multicolumn{5}{c}{NN.vs}\\ 
 \hline
 Leukemia    & 0.056 &  512 &  23 (4, 134)&   0.17 (0.14, 0.24)\\ 
 Breast 2 cl.& 0.337 &   88 &  23 (4, 110)&   0.24 (0.2, 0.31)\\ 
 Breast 3 cl.& 0.424 &    9 &  45 (6, 214)&   0.66 (0.61, 0.72)\\ 
 NCI 60      & 0.237 & 1718 & 880 (360, 1718)&   0.44 (0.34, 0.57)\\ 
 Adenocar.   & 0.181 & 9868 &  73 (8, 1324)&   0.13 (0.1, 0.18)\\ 
 Brain       & 0.194 & 1834 & 158 (52, 601)&   0.16 (0.12, 0.25)\\ 
 Colon       & 0.158 &    8 &   9 (4, 45)&   0.57 (0.45, 0.72)\\ 
 Lymphoma    & 0.04 &   15 &  15 (5, 39)&    0.5 (0.4, 0.6)\\ 
 Prostate    & 0.081 &    7 &   6 (3, 18)&   0.46 (0.39, 0.78)\\ 
 Srbct       & 0.031 &   11 &  17 (11, 33)&    0.7 (0.66, 0.85)\\ 
 \hline

 \end{tabular}
 }
 \end{center}
 \renewcommand{\baselinestretch}{0.2}\footnotesize\normalsize\footnotesize
 {
   $^*$Only two genes are selected from the complete data set; the values are the actual
   frequencies of those two genes.\\
   $^{\dagger}$\citet{shrunkenc} select 21 genes after visually inspecting 
   the plot of
   cross-validation error rate vs. amount of shrinkage and number of
   genes. Their procedure is hard to automate and thus it is very difficult to obtain estimates of the error
   rate of their procedure.\\
   $^{\ddagger}$\citet{shrunkenc} select 43 genes. The difference is likely due
   to differences in the random partitions for cross-validation. Repeating 100 times
   the gene selection process with the full data set the median, 1st quartile, and 3rd
   quartile of the number of selected genes are 13, 8, and 147.\\

 }
 \end{table}

\section{Discussion}

We have examined the performance of an approach for gene selection using random
forest, and compared it to alternative approaches. Our results, using both
simulated and real microarray data sets, show that this method of gene
selection accomplishes the proposed objectives.  Our method returns very small
sets of genes compared to two alternative variable selection methods, while
retaining predictive performance comparable to that of seven alternative
state-of-the-art methods.  Recently, \citet{BMA-selection} have proposed a
Bayesian model averaging (BMA) approach for gene selection; comparing the
results for the two common data sets between our study and theirs, in one case
(Leukemia) our procedure returns a much smaller set of genes (2 vs. 15),
whereas in another (Breast, 2 class) their BMA procedure returns 8 fewer genes
(14 vs. 6); our procedure does not require setting a limit in the maximum
number of relevant genes to be selected nor does it require to prespecify a
number of top ranked genes as relevant (the latter is nor required by the BMA
procedure either).  

Our method of gene selection will not return sets of genes
that are highly correlated, because they are redundant.  This method will be
most useful under two scenarios: a) when considering the design of diagnostic
tools, where having a small set of probes is often desirable; b) to help
understand the results from other gene selection approaches that return many
genes, so as to understand which ones of those genes have the largest signal to
noise ratio and could be used as surrogates for complex processes involving
many correlated genes. A backwards elimination method, precursor to the one
used here, has been already used to predict breast tumor type based on
chromosomic alterations \citep{SaraRF}.

We have also throughly examined the effects of changes in the
parameters of random forest (specifically $mtry$, $ntree$, $nodesize$)
and the variable selection algorithm ($se$, $fraction.dropped$).
Changes in these parameters have in most cases negligible effects,
suggesting that the default values are often good options, but we can
make some general recommendations. 
Time of execution of the code increases $\approx$ linearly with $ntree$.
Larger $ntree$ values lead to slightly more stable values of variable
importances, but for the data sets examined, $ntree = 2000$ or $ntree = 5000$
seem quite adequate, with further increases having negligible effects. The
change in $nodesize$ from 1 to 5 has negligible effects, and thus its default
setting of 1 is appropriate.  For the backwards elimination algorithm, the
parameter $fraction.dropped$ can be adjusted to modify the resolution of the
number of variable selected; smaller values of $fraction.dropped$ lead to finer
resolution in the examination of number of genes, but to slower execution of
the code.  Finally, the parameter $se$ has also minor effects on the results of
the backwards variable selection algorithm but a value of $se = 1$ leads to
slightly more stable results.

The final issue addressed in this paper is instability or multiplicity of the
selected sets of genes. From this point of view, the results are slightly
disappointing. But so are the results of the competing methods. And so are the
results of most examined methods so far with microarray data, as shown in
\citet{EinDor} and \citet{Michielis} and discussed throughly by
\citet{Somorjai2003} for classification and by \citet{pan-pnas} for the related
problem of the effect of threshold choice in gene selection.  However, and
except for the above cited papers and the review in \citet{Yo-azuaje}, this is
an issue that still seems largely ignored in the microarray literature. As
these papers and the statistical literature on variable selection
\citep[e.g.,][]{breiman-2-cultures, harrell-01} discusses, the causes of the
problem are small sample sizes and the extremely small ratio of samples to
variables (i.e., number of arrays to number of genes). Thus, we might need to
learn to live with the problem, and try to assess the stability and robustness
of our results by using a variety of gene selection features, and examining
whether there is a subset of features that tends to be repeatedly selected.
This concern is explicitly taken into account in our results, and facilities
for examining this problem are part of our R code.

The multiplicity problem, however, does not need to result in large
prediction errors.  This and other papers \citep{dudoit-dlda, pelora,
  simon.book, romualdi-03, bag-boost, Somorjai2003} show that very different
classifiers often lead to comparable and successful error rates with
a variety of microarray data sets. Thus, although improving prediction
rates is important \citep[specially if giving consideration to ROC
curves, and not just overall prediction error rates;][]{pepe-book},
when trying to address questions of biological mechanism or discover
therapeutic targets, probably a more challenging and relevant issue is
to identify sets of genes with biological relevance.

Two areas of future research are using random forest for the selection of
potentially large sets of genes that include correlated genes, and improving
the computational efficiency of these approaches; in the present work, we have
used parallelization of the ``embarrassingly parallelizable'' tasks using MPI
with the Rmpi and Snow packages \citep{Rmpi, snow} for R. In a broader context,
further work is warranted on the stability properties and biological relevance
of this and other gene-selection approaches, because the multiplicity problem
casts doubts on the biological interpretability of most results based on a
single run of one gene-selection approach.



\section{Conclusion}
The proposed method can be used for variable selection fulfilling the
objectives above: we can obtain very small sets of non-redundant genes while
preserving predictive accuracy. These results clearly indicate that the
proposed method can be profitably used with microarray data. Given its
performance, random forest and variable selection using random forest should
probably become part of the ``standard tool-box'' of methods for the analysis
of microarray data.

\section{Acknowledgements}


Most of the simulations and analyses were carried out in the Beowulf
cluster of the Bioinformatics unit at CNIO, financed by the RTICCC
from the FIS; J.~M.\ Vaquerizas provided help with the administration
of the cluster. A.\ Liaw provided discussion, unpublished manuscripts,
and code.  C.\ Lázaro-Perea provided many discussions and comments on
the ms. A.\ Sánchez provided comments on the ms.  I.\ Díaz showed
R.D.-U. the forest, or the trees, or both.  R.D.-U. partially
supported by the Ramón y Cajal program of the Spanish MEC (Ministry
of Education and Science); S.A.A. supported by project C.A.M.
GR/SAL/0219/2004; funding provided by project TIC2003-09331-C02-02 of
the Spanish MEC.

\bibliography{signatures2}
\bibliographystyle{bioinformatics}

\newpage

\end{document}